%% file: Clemente.Sundert.VPPC24.tex
\documentclass[conference]{IEEEtran}
\IEEEoverridecommandlockouts
\renewcommand{\baselinestretch}{0.97}
\usepackage[nolist,nohyperlinks]{acronym}
\input{Report/Packages.tex}%put packages in this document.
\input{Report/Acronyms.tex}


\usepackage[subtle]{savetrees}
\acrodef{GHGs}[GHGs]{greenhouse gas emissions}
\acrodef{LCA}[LCA]{life cycle assessment}
\acrodef{BEVs}[BEVs]{battery electric vehicles}
\acrodef{VIPV}[VIPV]{vehicle-integrated photovoltaic}
\renewcommand\figurename{Fig.}
\begin{document}

\title{A Framework to Estimate Life Cycle Emissions for Vehicle-Integrated Photovoltaic Systems %of Solar-powered Battery Electric Vehicles
\thanks{\noindent Maurizio Clemente, Mauro Salazar and Theo Hofman are with the Control Systems Technology section, Department of Mechanical Engineering, Eindhoven University of Technology (TU/e), Eindhoven, 5600 MB, The Netherlands.
E-mails: {\tt\footnotesize m.clemente@tue.nl}, {\tt\footnotesize m.r.u.salazar@tue.nl}, {\tt\footnotesize t.hofman@tue.nl} \newline
This paper was supported by the NEON research
project (project number 17628 of the Crossover program which
is (partly) financed by the Dutch Research Council (NWO)). }
}

\author{\IEEEauthorblockN{Maurizio Clemente, Luuk van Sundert, Mauro Salazar and Theo Hofman}
%	
% Accoding to template
%\author{\IEEEauthorblockN{Jorn van Kampen}
%	\IEEEauthorblockA{\textit{Control Systems Technology} \\
%		\textit{Eindhoven University of Technology}\\
%		Eindhoven, The Netherlands \\
%		j.h.e.v.kampen@tue.nl}
%	\and
%	\IEEEauthorblockN{Mauro Salazar}
%	\IEEEauthorblockA{\textit{Control Systems Technology} \\
%		\textit{Eindhoven University of Technology}\\
%		Eindhoven, The Netherlands \\
%		m.r.u.salazar@tue.nl}
%	\and
%	\IEEEauthorblockN{Theo Hofman}
%	\IEEEauthorblockA{\textit{Control Systems Technology} \\
%		\textit{Eindhoven University of Technology}\\
%		Eindhoven, The Netherlands \\
%		t.hofman@tue.nl}
%}

%\IEEEauthorblockA{\textit{Control Systems Technology} \\
%\textit{Eindhoven University of Technology}\\
%Eindhoven, The Netherlands \\
%j.h.e.v.kampen@tue.nl}
%\and
%\IEEEauthorblockN{Mauro Salazar}
%\IEEEauthorblockA{\textit{Control Systems Technology} \\
%	\textit{Eindhoven University of Technology}\\
%	Eindhoven, The Netherlands \\
%m.r.u.salazar@tue.nl}
}

\maketitle

\begin{abstract}
	This paper presents a framework to estimate the environmental impact of solar electric vehicles, accounting for the emissions caused by photovoltaic system production as well as vehicle use.
	We leverage a cradle-to-gate life cycle assessment to estimate the greenhouse gas emissions of the vehicle-integrated photovoltaic system, from the raw material extraction to the final panel assembly, including the effect of the electricity mix both at the factory location and in the country of use. %the vehicle's life cycle, considering both
	Furthermore, we modify an existing optimization framework for battery electric vehicles to optimally design a solar electric vehicle and estimate its energy consumption.
	We showcase our framework by analyzing a case study where the mono-crystalline silicon extraction and refinement processes occur in China, while the final assembly of the panel is in The Netherlands, generating 118 kg of $\mathrm{CO_2}$ equivalents per square meter of solar panel.
%	Moreover, we analyze the vehicle operations in different European countries, to assess if the additional emissions from the panel production result in a lower carbon footprint overall, comparing the operation-related emissions of a solar vehicle with those of a conventional battery electric vehicle, whereby both powertrains are designed to minimize energy consumption.
	The results suggest that it is generally beneficial to operate solar electric vehicles in countries with a high irradiation index.
	However, when the local electricity mix already displays a low carbon intensity, the additional emissions introduced by the panel are unnecessary, requiring a longer vehicle lifetime to reach an advantageous emission balance.

\end{abstract}


\input{Chapters/Introduction}

\input{Chapters/Methodology}

\input{Chapters/Results}

\input{Chapters/Conclusion}

\section*{Acknowledgment}
We thank Dr.~I.~New, Ir.~O.~J.~T.~Borsboom, Ir.~F.~Vehlhaber, Ir.~L.~Pedroso, and P.~Maharjan for proofreading this paper.
%Ir.~F.~Vehlhaber,
\bibliographystyle{IEEEtran}
\bibliography{main,SML_papers,references}

\end{document}

%% file: Report/Packages.tex
\usepackage{cite}
\usepackage{xcolor} % removed

\usepackage{mathtools} 
\usepackage{amsthm}

\usepackage{pgfplots}
\usepgfplotslibrary{groupplots}

\usepackage{hyperref}
\usepackage{amsmath}
\usepackage{amsfonts}
\usepackage{graphicx} % for pdf, bitmapped graphics files
\usepackage{amssymb}  % assumes amsmath package installed
\usepackage{pstricks,pst-plot,psfrag}
\usepackage{units}
\usepackage{float}
\usepackage{dsfont}
\usepackage{lettrine}
\usepackage{import}
\usepackage[acronym]{glossaries}

\usepackage{tikz}
\usetikzlibrary{backgrounds,calc,angles,quotes,arrows.meta,arrows,fit,shapes.geometric,shapes.multipart,positioning}
\usetikzlibrary{circuits.ee.IEC}
\usetikzlibrary{shapes.gates.logic.US}

%% file: Report/Acronyms.tex
\newacronym{acr:cvt}{CVT}{continuously variable transmission}
\newacronym{acr:CoG}{CoG}{center of gravity}
\newacronym{acr:CoP}{CoP}{center of pressure}
\newacronym{acr:CV}{CV}{constant-velocity}
\newacronym{acr:dp}{DP}{dynamic programming}
\newacronym{acr:DoF}{DoF}{degrees of freedom}
\newacronym{acr:ecms}{ECMS}{equivalent consumption minimization strategies}
\newacronym{acr:eltms}{ELTMS}{equivalent lap time minimization strategies}
\newacronym{acr:em}{EM}{electric motor}
\newacronym{acr:es2k}{ES2K}{Energy Storage to Kinetic}
\newacronym{acr:F1}{F1}{Formula 1}
\newacronym{acr:FIA}{FIA}{F\'{e}d\'{e}ration Internationale de l'Automobile}
\newacronym{acr:fgt}{FGT}{fixed-gear transmission}
\newacronym{acr:FD}{FD}{final drive}
\newacronym{acr:ice}{ICE}{internal combustion engine}
\newacronym{acr:k2es}{K2ES}{Kinetic to Energy Storage}
\newacronym{acr:mgu}{MGU}{motor generator unit}
\newacronym{acr:mguh}{MGU-H}{motor generator unit heat}
\newacronym{acr:mguk}{MGU-K}{motor generator unit kinetic}
\newacronym{acr:mpc}{MPC}{model predictive control}
\newacronym[description={energy management strategy}, \glslongpluralkey={energy management strategies},\glsshortpluralkey={EMSs}]{EMS}{EMS}{energy management strategy}%
\newacronym{acr:ODE}{ODE}{ordinary differential equation}
\newacronym{acr:pmp}{PMP}{Pontryagin's Minimum Principle}
\newacronym{acr:pu}{PU}{power unit}
\newacronym[description={powertrain operation}, \glslongpluralkey={powertrain operations},\glsshortpluralkey={POs}]{acr:PO}{PO}{powertrain operation}%
\newacronym{acr:socp}{SOCP}{second-order cone program}
\newacronym{acr:soe}{SoE}{state of energy}

%% file: Chapters/Introduction.tex
\section{Introduction}\label{sec:Introduction}

%\ac{BEVs} can provide a number of environmental benefits to our society: For instance, they can reduce the pollution in our cities thanks to the absence of tailpipe exhausts and help decrease global \ac{GHG} emissions.
%, if the energy used comes from renewable sources.
%While the elimination of local emissions of the vehicles may result in healthier cities, the electricity employed to power the vehicle still causes emissions where the energy is produced, shifting the problem upstream.
%The amount of carbon dioxide and other \ac{GHG}s produced to harvest electric energy depends on the country's electricity grid and, in case of a very dark mix, can lead to larger emissions from electric vehicles when compared to fossil-fuel-powered counterparts.
%the carbon intensity of the electric energy depends entirely on the country electricity mix, which can lead to larger emissions in some cases.
%The \ac{VIPV} technology offers the opportunity to generate clean energy directly on the vehicle, independent of the country's electricity mix, reducing the reliance on the local energy grid.
%However, when comparing the emissions of \ac{VIPV}s with standard \ac{BEVs}, it is crucial to take into account the additional emissions owing to the panel production.
%
%\ac{BEVs} can provide a number of environmental benefits to our society: For instance, they can reduce the pollution in our cities thanks to the absence of tailpipe exhausts and help decrease global \ac{GHG} emissions.
%, if the energy used comes from renewable sources.

\begin{figure}[t]
	\centering
	\includegraphics[width=0.85\columnwidth]{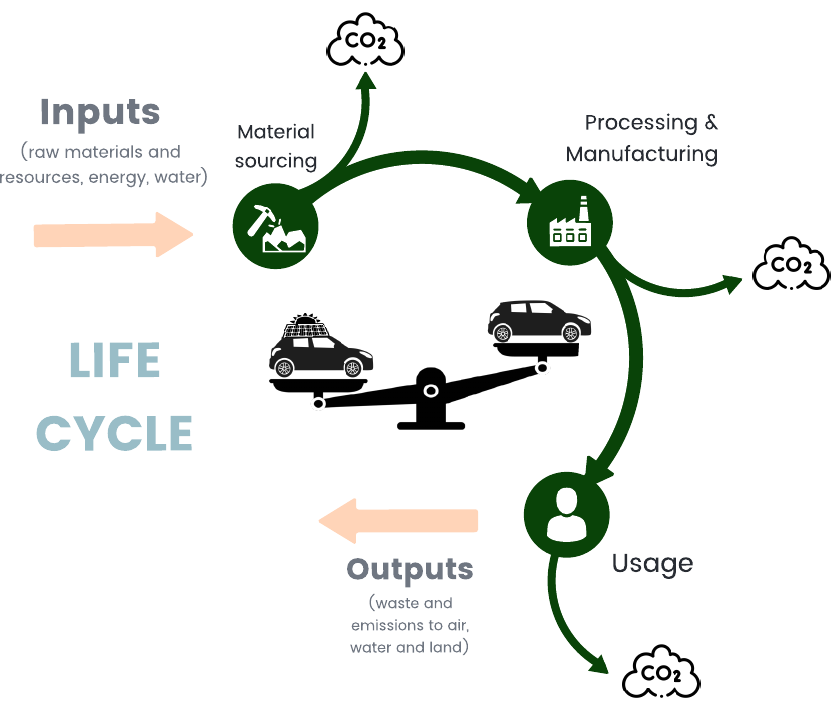}
	\caption{The cradle-to-gate life cycle assessment estimates the environmental impact (in terms of equivalent kilograms of $\mathrm{CO_2}$) from the raw material extraction to the final panel assembly. In our framework, we also include the vehicle use phase emissions which depend on the powertrain design and its energy consumption.}
	\label{fig:Meth}
\end{figure}

\begin{figure}[t]
	\centering
	\includegraphics[width=0.9\columnwidth]{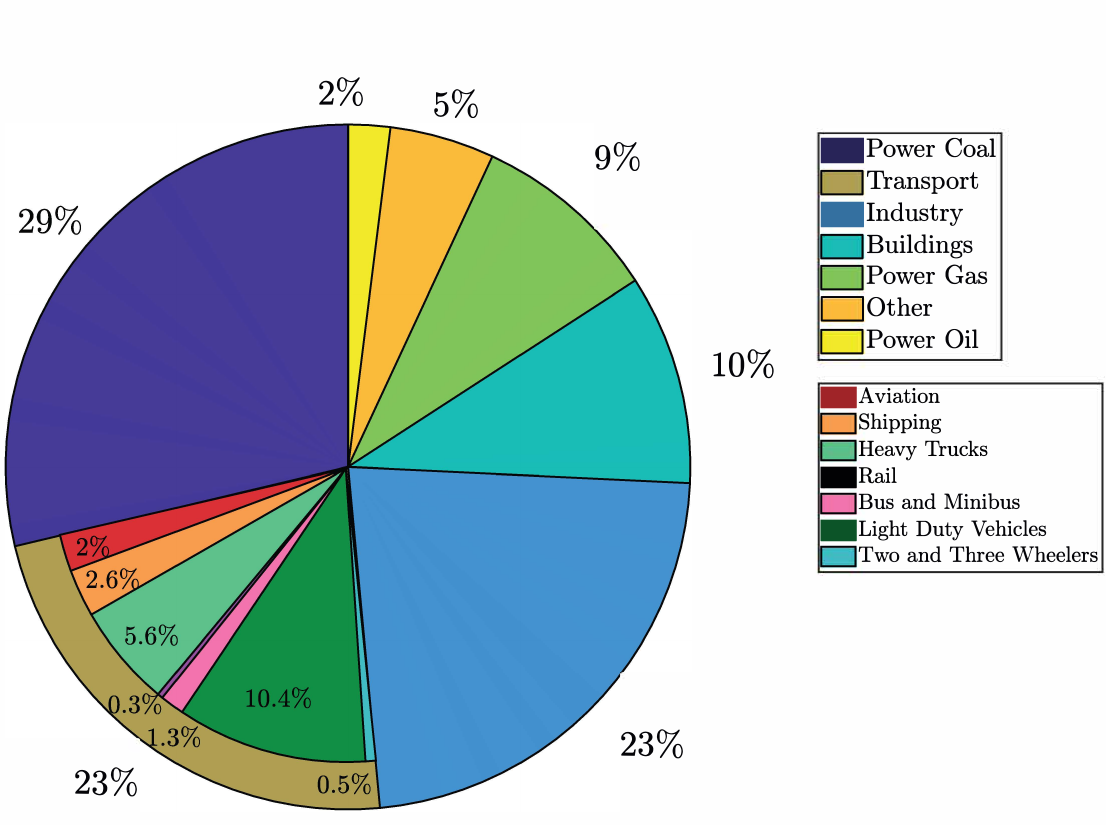}
	\caption{Global emissions by sector and sub-sector. Adapted from data in \cite{IEA2020,IEA_global_21}.}
	\label{fig:Header}
\end{figure}

The transportation sector alone currently contributes up to 23\% of the global \ac{GHGs}.
In the effort to reduce its environmental impact, we face an unprecedentedly difficult and multidisciplinary challenge.
Among the transportation types analyzed in \cite{IEA2020}, light-duty vehicles have the highest contribution to \ac{GHGs}.
For this reason, many researchers have proposed new ideas and solutions to mitigate the environmental impact of road vehicles without severely compromising the modern paradigm of mobility.
For instance, in recent years, the share of highly energy-efficient electric vehicles on the market has increased exponentially.
Although the abolition of tailpipe emissions contributes to reducing local pollution, \ac{BEVs} still require electric energy to be operated.
If the energy comes from non-renewable sources, the risk is to shift the emissions upstream~\cite{AR6} in the energy harvesting line, without tackling the problem directly (Fig.~\ref{fig:Meth}).
As a result, the global \ac{GHGs} caused by light-duty vehicles could be re-labeled under the Industry and Power sectors (Fig.~\ref{fig:Header}).
In recent years, the amount of research and diffusion of solar cells has enabled higher energy generation densities and lower costs~\cite{CentenoBrito2021UrbanPhotovoltaics}, paving the way for innovative applications such as \ac{VIPV} systems (i.e., solar electric vehicles). %on the market
By installing a clean energy generation system directly on the vehicles, it is possible to address the emission shift issue while at the same time substantially reducing the reliance on the local energy grid~\cite{Celik2023SustainableApplications} of conventional \ac{BEVs}.
However, solar panels are manufactured starting from silicon extracted in mines~\cite{SpecialChains} and are often a source of neglected \ac{GHGs}.
Hence, it becomes utterly important to predict the emission generated during panel manufacturing to understand whether it is more sustainable to generate energy on-board or use traditional \ac{BEVs} and rely on the local electricity mix.
Against this backdrop, this paper presents a framework to estimate the environmental cost of solar vehicles and compare it with traditional \ac{BEVs}, assessing under which conditions it is beneficial to adopt \ac{VIPV} systems.
We leverage a cradle-to-gate \ac{LCA} of a mono-crystalline silicon panel, while the lifetime energy consumption and, in turn, the operation-related emissions are computed via a suitably modified convex optimization framework taking into account panel, motor, and battery sizes.

% prompting a possible application of this technology.
%out of the ransportation types analyzed in Fig.~\ref{fig:Header}, light-duty vehicles have the highest contribution to \ac{GHGs}, prompting a possible application of this technology.

%footprint-aware design framework for solar electric vehicles, 
%optimizing the total amount of \ac{GHG} emissions during the vehicle's lifetime 
%   and operations-related emissions linked to the electricity utilization from the local energy grid.
%Although the main contributor to the emissions is the battery~\cite{AichbergerJungmeier2020,ChordiaNordeloefEtAl2021},
% and potentially 
%mono-crystalline silicon

\emph{Related Literature:}
This paper pertains to two main research lines: The first one is the \ac{LCA} of photovoltaic systems. 
%Traditionally, \ac{LCA} is employed to evaluate the overall impact of an item on a specific quantity (e.g. monetary or environmental cost), during its whole life cycle~\cite{ISO2020}.
According to the ISO14040 standard~\cite{ISOFramework}, the \ac{LCA} methodology is a systematic analysis of the overall impact of an item on a specific quantity (e.g., monetary or environmental cost), during its entire life cycle, usually divided into material extraction, components production, distribution, use, and end-of-life.
We consider a mono-crystalline silicon panel cradle-to-gate \ac{LCA}, accounting for the environmental impact from the raw material extraction to the final assembly.
Many studies, summarized in \cite{Muteri2020ReviewPanels}, have explored the emissions generated during the manufacturing process of photovoltaic systems, leading to a plethora of models~\cite{PVPSTaskLifeSustainability,HsuODonoughueEtAl2012}.
However, most of the results in these studies need to be updated due to recent developments in technology.
Specifically, the energy usage in manufacturing has almost halved~\cite{SpecialChains}, resulting in a significantly lower environmental impact of manufacturing, as a large part of the emissions owing to photovoltaic panel production is dependent on the carbon intensity of the electricity mix. 
Moreover, most studies employ the energy produced by the panel in kilowatt-hours as a functional unit~\cite{ISOFramework,Curran2006USPractice}, a quantity that is influenced by different conditions, often requiring harmonization. %like temperature,
Although the environmental impact of photovoltaic panels has been abundantly studied, to the best of the authors' knowledge, the impact has never been compared with the change in energy consumption owing to the panel, in the context of analyzing the life cycle emissions of \ac{VIPV} systems as opposed to conventional \ac{BEVs}.

The second research stream concerns solar electric powertrain design.
As electric vehicle technology matured in recent years, powertrain design has seen huge developments, culminating in a broad spectrum of methods and applications~\cite{GuzzellaSciarretta2007,SilvasHofmanEtAl2016,WangZhouEtAl2022}.
The \ac{VIPV} technology is currently being explored in Europe through the SolarMoves project, commissioned by the Department for Mobility and Transport of the European Commission, quantitatively assessing solar electricity generation on vehicle bodies and its impact on the future charging infrastructure in Europe.
Whilst powertrain design for hybrid~\cite{BorsboomFahdzyanaEtAl2021,CarlosDaSilvaKefsiEtAl2023} and conventional battery electric vehicles~\cite{BorsboomSalazarEtAl2022,ClementeSalazarEtAl2024} has been explored by many authors, the optimal design for a solar electric vehicle has been limited to university teams and competition, apart from rare exceptions~\cite{Lightyear2022}.

%Nevertheless, none of these methods focused on the vehicle lifetime emissions.

%\textbf{	I would have thought that was necessary.system can be clarified in the statement of contribution.
%	The use phase is also something that belongs to the LCA approach, which is not clear in the text.}
\emph{Statement of Contribution:}
In this study, we provide an \ac{LCA} framework to estimate the emissions generated by solar electric vehicles. 
The manufacturing process of the photovoltaic panel is investigated through an \ac{LCA} approach, from the raw material extraction until its final assembly (cradle-to-gate analysis), while the emissions generated during the operation phase derive from the energy consumption of an optimally designed powertrain.
For this reason, we opportunely modified an existing framework~\cite{ClementeSalazarEtAl2022} for \ac{BEVs} to take into account the energy produced by the panel.
Furthermore, in our \ac{LCA} we adopt square meters as opposed to kilowatt-hour as a functional unit.
This way, the energy produced by the panel is not dependent on temperature and can be easily scaled in a powertrain design optimization framework.
Finally, we analyze the impact of the panel production location, country of use, solar panel dimension, and vehicle lifetime on the trade-off, deriving useful insights on the condition in which the \ac{VIPV} technology offers advantage over conventional \ac{BEVs}.

\emph{Organization:}
The remainder of this study is structured as follows: In Section \ref{sec:Methodology} we instantiate our framework, presenting the photovoltaic panel \ac{LCA} and the vehicle's powertrain and optimal energy consumption model.
In Section \ref{sec:Results} we exemplify our methodology in a realistic case study and discuss the main findings.
The conclusions of the study are summarized in Section \ref{sec: Conclusion}, together with an outline of future research.

%% file: Chapters/Methodology.tex
\section{Methodology} \label{sec:Methodology}

In this section, we illustrate our framework in detail.
In particular, Section~\ref{sec:C2GLCA} defines the photovoltaic panel \ac{LCA} methodology to estimate the panel manufacturing environmental impact $I_{\mathrm{p}}$, capturing the manufacturing process and the system boundaries.
Section~\ref{sec:ORE} treats the operations-related emissions generated during the vehicle's lifetime $I_\mathrm{o}$, presenting the equations introduced to adapt the original \ac{BEVs} powertrain optimization tool to \ac{VIPV} systems.
Hence, the overall environmental impact $I$ can be expressed as 
\begin{equation*}\label{eq:tot}
	I = I_{\mathrm{p}} + I_{\mathrm{o}},
\end{equation*}
where $I_{\mathrm{p}}$ is null for conventional \ac{BEVs}.
Finally, in Section~\ref{sec:DISC} we discuss assumptions and limitations of our approach.

\subsection{Cradle-to-gate Life Cycle Assessment}\label{sec:C2GLCA}

\begin{figure} 
	\centering
	\includegraphics[width=0.85\linewidth]{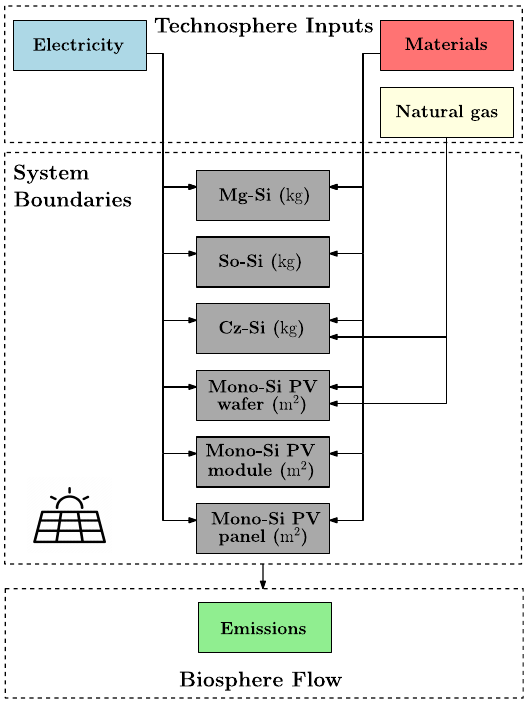}
	\caption{Diagram including the system boundaries and the processes leading to the final assembly of the panel from the raw material extraction, considering energy and materials flows. Biosphere flows are direct emissions from a process (downstream), while technosphere flows are required for the production of the end product (upstream). The complete inventory is available in the database~\cite{ClementevanSundert2024}.}\label{fig:process}
\end{figure}

The \ac{LCA} consists of four different phases: goal and scope definition phase, life cycle inventory phase, impact assessment phase, and interpretation phase.
The first phase, as the name suggests, establishes the purpose of the study, together with the functional unit through which the impact of the product is quantified~\cite{ArzoumanidisDEusanioEtAl2020}.
The life cycle inventory details the inflow and outflow of materials, energy, waste products, and emissions of the process within the system boundaries, whereas the life cycle impact assessment quantifies the key performance indicator selected according to the functional unit.
Finally, the interpretation of the results reconnects to the purpose of the study, allowing fruitful considerations.

In this paper, we examine the manufacturing of a photovoltaic panel, analyzing the environmental impact in terms of global warming potential, expressed as equivalent kilograms of $\mathrm{CO_2}$ generated in the production process. 
Out of the several existent photovoltaic panel technologies, we focus our attention on the emission generated by mono-crystalline silicon panels.
This choice can be ascribed to their diffusion on the \ac{VIPV} systems industry, where they are deemed the best compromise considering performance, costs, stability, and availability~\cite{CentenoBrito2021UrbanPhotovoltaics}.

% whereby we consider the materials and energy flows from the mineral extraction to the final product assembly.
In Fig.~\ref{fig:process} we define the system and identify the processes that lead to the final photovoltaic panel assembly from the raw material extraction, accounting for all the flows of materials, energy, waste products, and emissions inside and outside the system boundaries.
In particular, we subdivide the panel manufacturing process into six different production steps: metallurgical-silicon (Mg-Si) production, solar grade silicon (So-Si) production, mono-crystalline Czochralski silicon (Cz-Si) production, wafer production, cell production and panel production. %The first three silicon purification steps are also referred to as mono-crystalline silicon production.
We refer all the flows to the manufacturing of one square meter of a photovoltaic panel, favoring this functional unit over the conventionally adapted kilowatt hour of energy produced.
As opposed to the kWh, this choice makes it easier to account for the product \ac{GHGs}, independently of the operative conditions (e.g., temperature), allowing for easier integration into the powertrain design optimization routine.
 
%In line with the conventional standards adopted~\cite{Curran2006USPractice}, we divide the inventory flows in two: biosphere and techno-sphere. Biosphere flows are direct emissions from a process (downstream), while Techno-sphere flows are required for the production of the end product (upstream).
We derive the data from the Ecoinvent 3.9database~\cite{Frischknecht2005TheFramework,WernetBauerEtAl2016} and build the model in the ``Activity Browser"~\cite{Steubing2020TheFramework} software, whilst leveraging an interface with the software ``Brightway2"~\cite{Mutel2017Brightway:Assessment} to efficiently change parameters, enabling large flexibility in running many different supply chain scenarios.
Specifically, we account for the electricity mix and transportation inputs dependency on the manufacturing location, to explore their impact on the \ac{GHGs}. 
Conversely, the other inputs are not dependent on the location, due to the lack of information about the material, energy, and environmental efficiencies in the database.
%ecoinvent 3.9.1 cutoff ecoSpold02 database.
Thereafter, we translate emissions and resource extractions into environmental impact scores employing characterization factors in the life cycle inventory assessment phase.
Two commonly adopted ways to derive characterization factors are midpoint and endpoint level indicators~\cite{Huijbregts2017ReCiPe2016:Level}: midpoint indicators focus on a single environmental problem, such as climate change or acidification. Endpoint indicators show the environmental impact on higher aggregation levels, such as ecosystem damage or human health.
In this study, we adopt the 2016 ``ReCipe midpoint method"~\cite{Huijbregts2017ReCiPe2016:Level} to compute the \ac{GHGs}.
For the purpose of this study, we only consider the midpoint indicators for the climate change category, leaving the others (photo-chemical ozone formation, terrestrial acidification, freshwater eutrophication, land use and fossil resource scarcity) for potential extensions of the framework in the future, allowing for highlighting potential trade-offs.
%by the use of different databases: The IEA considers ``UVEK LCI data DQRv2:2018" database, based on Ecoinvent 2.2 while we used Ecoinvent 3.9.
%Furthermore, the report uses a cradle-to-grave approach as opposed to our cradle-to-gate approach.
%This implies that the end of life phase of the PV system and the cleaning and maintenance during lifetime is also taken into account in the AEI paper. This could explain the slightly lower GHG emissions in the Panel production step.
%The life-cycle emissions are 15.38\% higher than the one computed by IEA \cite{SolarIEA}.
%However, we can attribute this discrepancies to the use of different databases~\cite{Lc-inventories}, and the system boundaries: IEA is using cradle-to-grave as opposed to cradle-to-gate.
%Furthermore, we grouped the Mg-Si, So-Si, and Cz-Si contributions under Mono-Si production, to compare it with their process subdivision.
%\begin{equation}\label{equation2}
%	M_{\mathrm{GHG}}=\frac{W}{I\cdot \eta \cdot P \cdot t_{\mathrm{lt}} \cdot A,}
%\end{equation}
%where I is the irradiation in $\frac{kWh}{m^2}$ per year, $\eta$ the lifetime average module efficiency, P the performance ratio, $t_{\mathrm{lt}}$ the system lifetime in years, W the GHG emitted over the lifetime of the PV system in gram $CO_2$ equivalent and A the total module area in $m^2$.

\subsection{Operations-related Emissions}\label{sec:ORE}

\begin{figure} 
	\centering
	\includegraphics[width=\columnwidth]{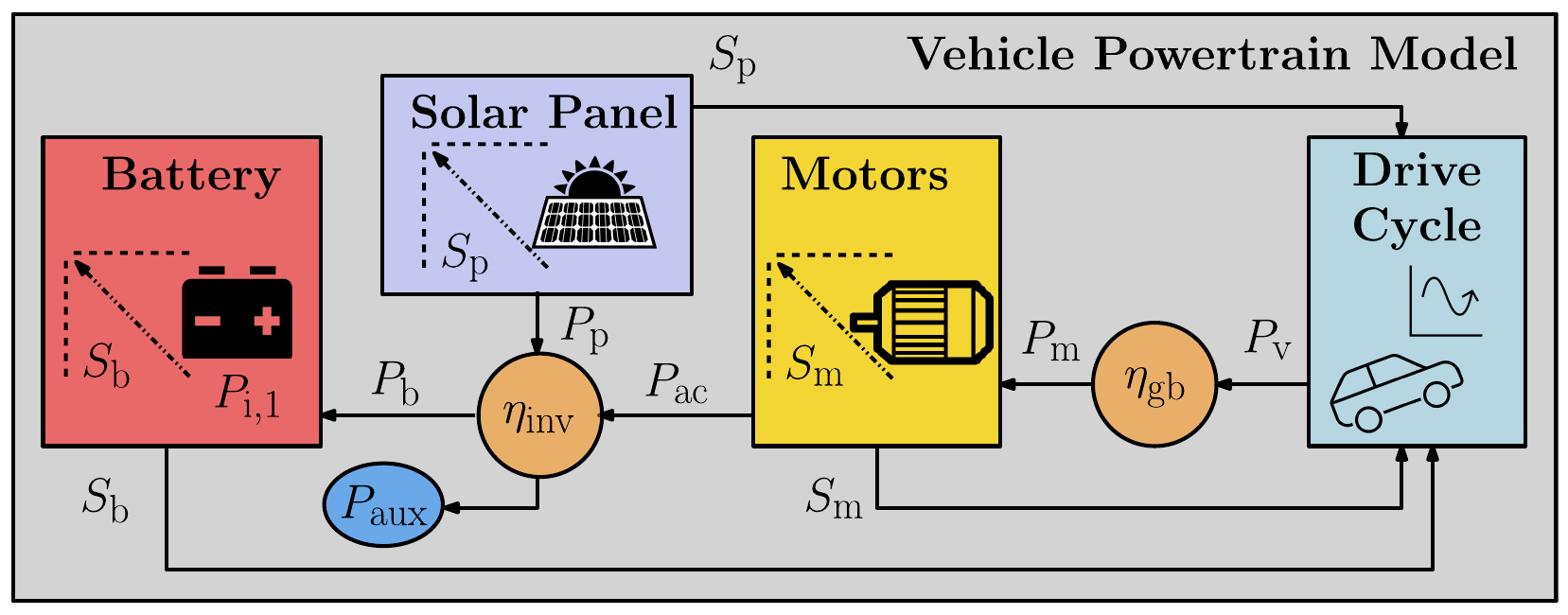}
	\caption{Vehicle powertrain and energy consumption model, modified from \cite{ClementeSalazarEtAl2022}.}\label{fig:ptmodel}
\end{figure}

In order to estimate the lifetime operations-related emissions, we encompass the influence of the distance-specific vehicle energy consumption $F_{\mathrm{v}}$, as well as the carbon intensity of the electricity mix in the country of usage $c$, and the vehicle lifetime $L$
\begin{equation*}\label{eq:Io}
	I_{\mathrm{o}} = F_{\mathrm{v}} \cdot c \cdot L.
\end{equation*}
We determine the vehicle energy consumption by modifying a previously developed model~\cite{ClementeSalazarEtAl2022} to jointly optimize the photovoltaic panel dimension $S_\mathrm{p}$ with the battery and motor sizing $S_\mathrm{b}$ and $S_\mathrm{m}$, respectively.
To this end, we introduce new equations to model the behavior of the solar panel, and we modify others to account for its impact on the rest of the powertrain (Fig.~\ref{fig:ptmodel}).
Furthermore, we include constraints on the vehicle performance to ensure that the design meets the minimum range $\underline{d}_\mathrm{r}$, top speed $\underline{v}_\mathrm{t}$, and acceleration time (0-100 km/h) $\overline{t}_\mathrm{a}$.
Finally, we retain the convexity of the formulation by applying lossless relaxations to preserve the convex properties of global optimality of the solution and convergence in polynomial time.
We consider the solar panel as a power generator whose output depends on the panel size and the average daily horizontal irradiation~\cite{GlobalAtlas},
\begin{equation}\label{eq:Sp}
	P_{\mathrm{p}} = S_{\mathrm{p}} \cdot \overline{P}_{\mathrm{c}} \cdot k_{\mathrm{HI}},
\end{equation}
where $P_{\mathrm{p}}$ is the power produced by the panel, $S_{\mathrm{p}}$ is the number of solar cells, $\overline{P}_{\mathrm{c}}$ is the maximum power of a solar cell, and $k_{\mathrm{HI}}$ is the horizontal irradiation coefficient.
Due to physical constraints, the maximum area available on the vehicle $A_{\mathrm{a}}$ determines the dimensions of the panel following the constraint
\begin{equation}\label{eq:A_max}
	S_{\mathrm{p}} \cdot A_{\mathrm{c}} \leq A_{\mathrm{a}},
\end{equation}
with $A_{\mathrm{c}}$ representing the area of a single photovoltaic cell.
Furthermore, we consider the additional weight introduced by the panel by adding a term to the vehicle mass equation.
Hence, the vehicle mass $m$ consists of several contributions: the glider mass $m_{\mathrm{g}}$, the driver mass $m_{\mathrm{d}}$, the payload mass $m_{\mathrm{pl}}$, and the motor, battery, and solar cell mass, computed by scaling the reference masses $m_{\mathrm{m,o}}$, $m_{\mathrm{b,o}}$, and $m_{\mathrm{c,o}}$ respectively,
\begin{equation}\label{eq:mass}
	m = m_{\mathrm{g}} + m_{\mathrm{d}} + m_{\mathrm{pl}} + m_{\mathrm{m,o}} \cdot S_{\mathrm{m}} + m_{\mathrm{b,o}} \cdot S_{\mathrm{b}}  + m_{\mathrm{c,o}} \cdot S_{\mathrm{p}}.
\end{equation}
Finally, the battery power $P_{\mathrm{b}}$ can be found starting from the motor input power $P_{\mathrm{ac}}$ by considering the inverter efficiency $\eta_\mathrm{inv}$, power consumption of auxiliary systems $P_{\mathrm{aux}}$ (heating, air conditioning, lights, etc.), and the power generated by the panel $P_{\mathrm{p}}$. 
Thus we can write the power bus equation as
\begin{align*}
	%	\label{eq:inverter}
	P_{\mathrm{b}} = 
	\begin{cases}
		 \frac{ P_{\mathrm{ac}} - P_{\mathrm{p}} + P_{\mathrm{aux}} }{\eta_\mathrm{inv}} \quad & \text{if } P_{\mathrm{ac}} \geq 0 \\
		 \eta_\mathrm{inv} \left( P_{\mathrm{ac}} - P_{\mathrm{p}} + P_{\mathrm{aux}} \right)\quad & \text{if } P_{\mathrm{ac}} < 0
	\end{cases},
\end{align*}
that can be written as 
\begin{equation}
	\label{eq:Cinverter}
	P_{\mathrm{b}} \geq \frac{ P_{\mathrm{ac}} - P_{\mathrm{p}} + P_{\mathrm{aux}} }{\eta_\mathrm{inv}},
\end{equation}
\begin{equation}
	\label{eq:Dinverter}
	P_{\mathrm{b}} \geq \eta_\mathrm{inv} \left( P_{\mathrm{ac}} - P_{\mathrm{p}} + P_{\mathrm{aux}} \right),
\end{equation}
following a lossless epigraphic relaxation of the constraint~\cite{Boyd2007}.
Due to the particular problem structure, the constraint will always hold with equality for the optimal solution.
In fact, assuming any value higher than the strict necessary would be sub-optimal as it entails a higher energy consumption.
We can formulate the optimization problem by introducing Equations~\eqref{eq:Sp} and \eqref{eq:A_max} in \textit{Problem 1} from \cite{ClementeSalazarEtAl2022} and replacing the original mass (4), and inverter equations (17) and (18), with Equations \eqref{eq:mass}, \eqref{eq:Cinverter}, and \eqref{eq:Dinverter} in this paper, respectively.
Ultimately, we minimize the energy consumption for a single vehicle (tailored design), converging rapidly to the global optimal solution with standard algorithms.
%and parametric values for solar radiation and vehicle lifetime, enabling the analysis and comparison of different scenarios.

 \subsection{Discussion}\label{sec:DISC}

A few comments are in order. 
First, we focus our attention specifically on the photovoltaic system. The emissions related to the production of other powertrain components are outside the scope of this paper and will be analyzed in future work.
Second, we are focusing only on a cradle-to-gate analysis for the panel manufacturing, not accounting for end-of-life and recycling in our analysis. However, according to \cite{HeldIlg2011}, this is a conservative assumption as the potential environmental benefits from material recycling and energetic recovery outweigh the negative impacts of the recycling process and therefore would lead to a reduction of the environmental profile.
Furthermore, we assume that the panel contributes only during vehicle operations, neglecting the energy produced when the vehicle is parked.
However, we average the energy produced based on the daily light hours in the country.
Finally, although the inventory data may be affected by uncertainties or change over time, the same proposed methodology can still be applied.

%The extension of the framework to overcome this assumption requires further research on the drivers habits and it varies a lot from one country to another.
%First, our model does not include a specific panel inverter due to lack of data. We considered most of the ones available on the market have performance in line with a motor inverter

%First, the model 
%Including this could have a increasing effect on the manufacturing phase emissions.
 
%Second, the location dependent LCA is limited to energy mix and changing transportation emissions. Other regional effects such as increases in efficiency thanks to scale advantages are not taken into account.
% In our use case this might effect the GHG emissions of supply chain 1. 
%  
%  Since, the Xinjiang region owns the largest Mono-Si manufacturing plant in the world \cite{SpecialChains}. 

 %The cleanest area in China with PV manufacturing sites is Sichuan.
 %Transportation 

%Batter
%Finally,  due to the limited 
%In a, the 
%for a 
%we compare it to savings  owing to the panel production

%% file: Chapters/Results.tex
\section{Results}\label{sec:Results}

\begin{figure}[t]
	\centering
	\includegraphics[width=0.9\columnwidth]{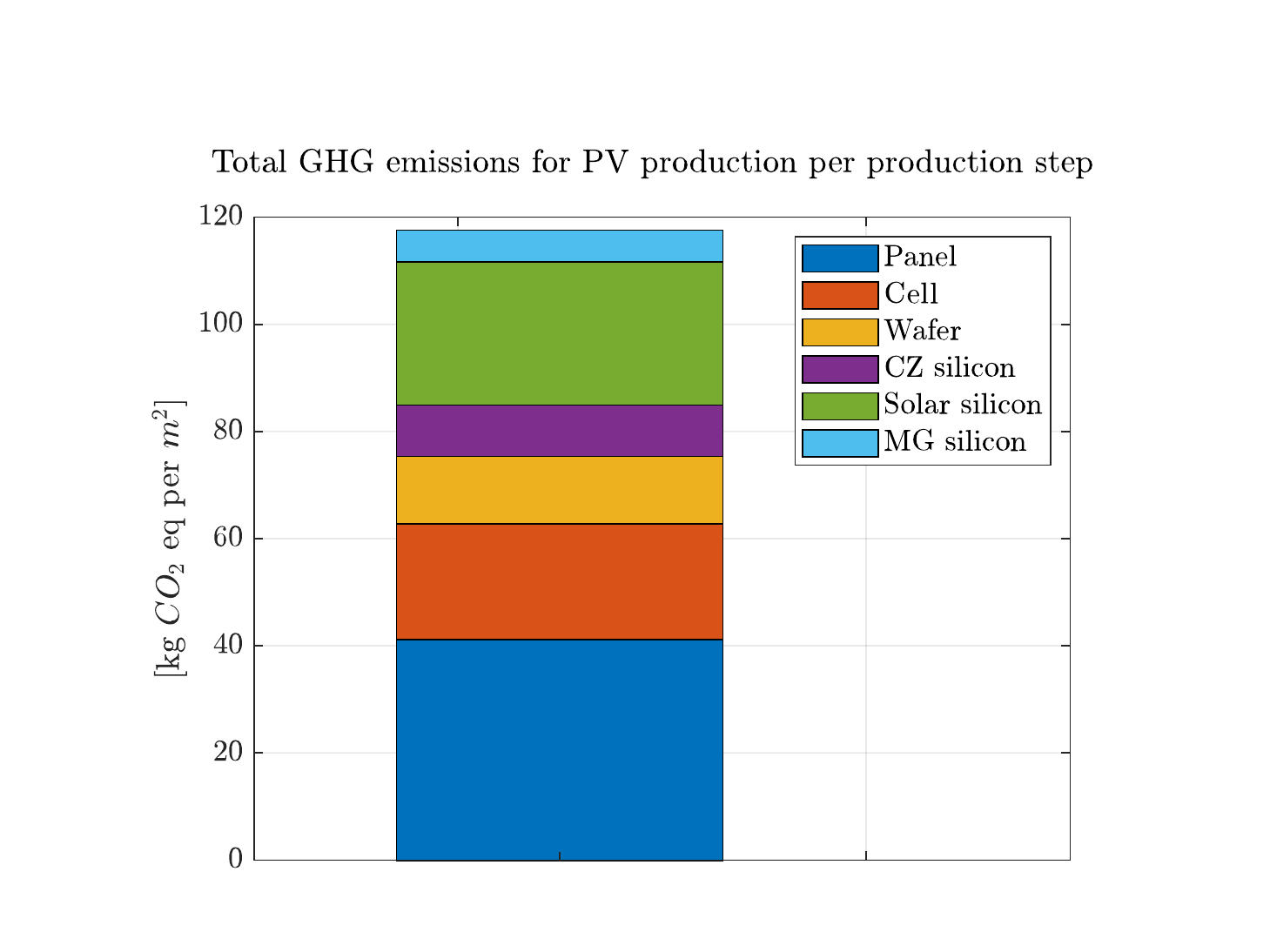}
	\caption{Single manufacturing processes contributions to the total \ac{GHGs} per square meter of photovoltaic system production.}
	\label{fig:bar}
\end{figure}

\begin{figure*}[t]
	\begin{minipage}{0.49\linewidth}
		\centering
		\includegraphics[width=0.9\linewidth]{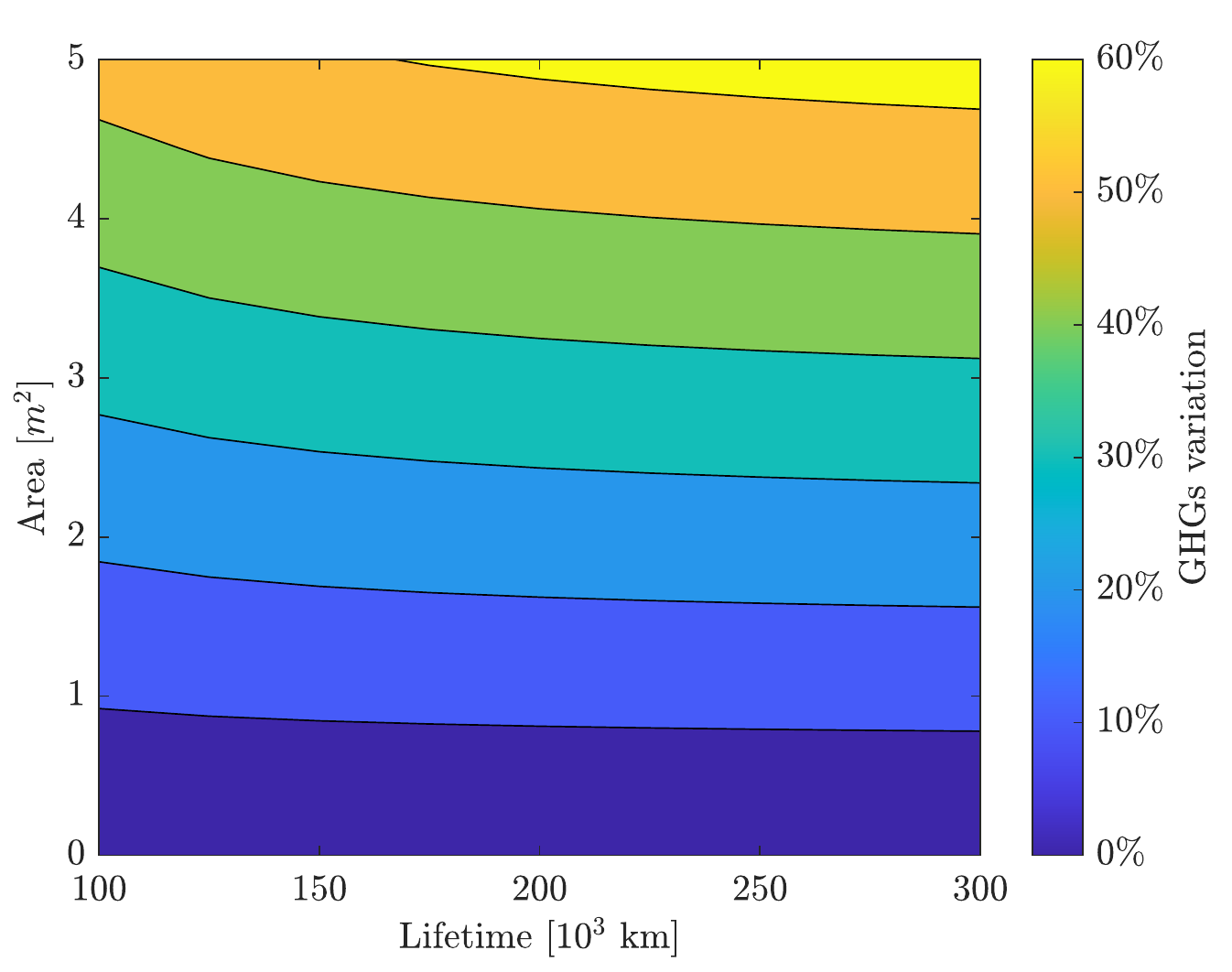}
		\caption{Relative difference in \ac{GHGs} between \ac{VIPV} systems and \ac{BEVs} operated in The Netherlands as a function of the installed panel area and vehicle lifetime.}
		\label{fig:NLc}
	\end{minipage}\hfill
	\begin{minipage}{0.49\linewidth}
		\centering
		\includegraphics[width=0.90\linewidth]{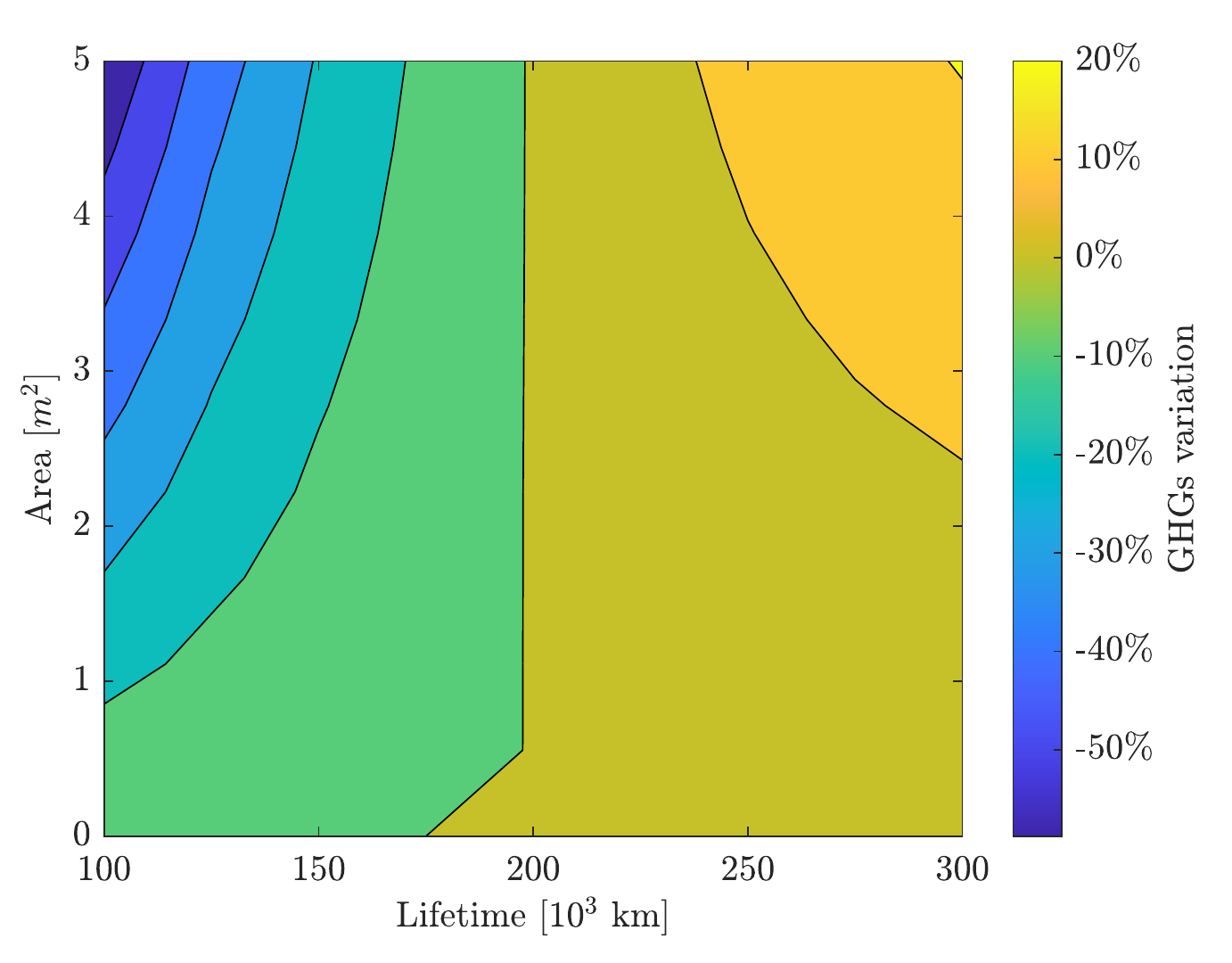}
		\caption{Relative difference in \ac{GHGs} between \ac{VIPV} systems and \ac{BEVs} operated in Sweden as a function of the installed panel area and vehicle lifetime.}
		\label{fig:SWc}
	\end{minipage}
\end{figure*}

In this section, we demonstrate our framework on a case study considering a \ac{VIPV} panel production line where the silicon is refined in China and the final panel assembly takes place in The Netherlands.
We identified China as the most suitable location to select for silicon processing since 79\% of the mono-crystalline silicon production is located in China, with 42\% alone in the Xinjiang region~\cite{SpecialChains}.
For our analysis, we consider a photovoltaic panel based on European solar cell data, reported in Table~\ref{tab:cel_par}, and installed on a city car, optimally designed to meet the performance in Table~\ref{tab:Minimum_Performance Parameters}, starting from the design parameters in Table~\ref{tab:vpar}.
We estimate the operation-related emissions based on the energy consumption to drive the Class 3 Worldwide harmonized Light-duty vehicles Test Procedure (WLTP).
We discretize the problem using the Euler forward method with a sampling time of 1 s, parsing it with YALMIP~\cite{Loefberg2004} and solving it to global optimality with MOSEK~\cite{ApS2017}, in approximately 2 s.

% Panel Data
\begin{table}[b]
	\centering
	\caption{Mono-crystalline silicon solar cell characteristics~\cite{Monocrystallineoptional}.}\label{tab:cel_par}
	\begin{tabular}{cc} \hline
		\textbf{Parameter} & \textbf{Value} \\ \hline
		$m_{\mathrm{c,o}}$& 0.28 kg \\ 
%		$V_{\mathrm{c}}$ & 0.62~\unit{V} \\ \hline
%		$i_{\mathrm{c}}$  & 7.88~\unit{A} \\ \hline
		$\overline{P}_{\mathrm{c}}$ & 4.88 W \\ 
		$A_{\mathrm{c}}$ & 0.0243 m$^2$ \\ 
		$\underline{k}_{\mathrm{HI}}$ & 0 \\ 
		$\overline{k}_{\mathrm{HI}}$ & 1 \\ 
		%		Thickness  $d_{\mathrm{c}}$ & 170 & \unit{$\mu m$} \\ 
		\hline
	\end{tabular}
\end{table}
% Performance Parameter Table
\begin{table}[b]
	\centering
	\caption{Required Performance Parameters.}
	\begin{tabular}{cc} \hline
		\label{tab:Minimum_Performance Parameters}
		\textbf{Parameter}  & \textbf{Value} \\ \hline
		$\underline{d}_\mathrm{r}$ & 200 km \\ 
		$\underline{v}_\mathrm{t}$ & 130 km/h \\ 
		$\overline{t}_\mathrm{a}$ & 15 s \\ 
		$P_{\mathrm{aux}}$ & 500 W \\ 
		\hline
	\end{tabular}
\end{table}
%The thickness of the module is assumed 170 $\mu m$ and has 60 individual cells of 243 $cm^2$, comprising a panel of 1.6 $m^2$. 
% Vehicle Parameter Table
\begin{table}[b]
	\centering
	\caption{Vehicle Parameters.}
	\label{tab:vpar}
	\begin{tabular}{l c}\hline
		\textbf{Parameter}  &   \textbf{Value} \\
		\hline
%		$A_{\mathrm{f}}$ & 2.38 & \unit{$m^2$}\\
%		$r_{\mathrm{w}}$  & 0.3498 &\unit{$m$}  \\
%		$c_{\mathrm{d}}$  & 0.29  & $-$ \\
%		$c_{\mathrm{r}}$  & 0.01  & $-$ \\
		$\eta_{\mathrm{inv}}$  & 0.96 \\
%		$r_{\mathrm{b}}$  & 1 & $-$  \\
		$m_{\mathrm{g}}$  & 850 kg  \\
		$ {m}_{\mathrm{d}} $  & 85 kg  \\
		$ {A}_{\mathrm{a}} $  & 3 m$^2$  \\
		\hline
	\end{tabular}
\end{table}

% 138.8~
The \ac{GHGs} of the considered case study reach 118 kg of $\mathrm{CO_2}$ equivalents per m$^2$ of PV panel.
This result is in line with the estimated emissions in the case studies proposed by other authors.
In a recent report from the IEA~\cite{FactIEA-PVPS}, where silicon production is staged in China, the estimated emissions for a kilowatt-hour of photovoltaic capacity is 27.0 kg of $\mathrm{CO_2}$ equivalents.
After converting our results (in square meters) with the harmonization methodology presented in \cite{HsuODonoughueEtAl2012}, we found a value of 23.4 kg of $\mathrm{CO_2}$ equivalents.
We can attribute the 13.3\% difference to the use of different databases~\cite{Lc-inventories} and system boundaries (cradle-to-grave as opposed to cradle-to-gate).
However, when comparing our model to a broader range of photovoltaic panel \ac{LCA}s~\cite{Muteri2020ReviewPanels}, the emissions estimations of our model turn out to be below average.
This result can be ascribed to the fact that all the papers included in the study were published prior to 2018, hence using less recent inventory data.
In fact, the electricity needed in manufacturing has more than halved in recent years~\cite{SpecialChains} thanks to more efficient use of energy and materials in the supply chain.
As can be observed in Fig.~\ref{fig:bar}, the main contribution to the manufacturing emissions is the panel framing, as it involves aluminum production, which is rather energy intense.
Another major contribution is solar-grade glass production, which requires large quantities of electrical energy and occurs in a country with a carbon-intense electricity mix.
Finally, transportation only plays a small role, accounting for just 0.81\% of the emissions.

\begin{figure}[t]
	\centering
	\includegraphics[width=0.9\columnwidth]{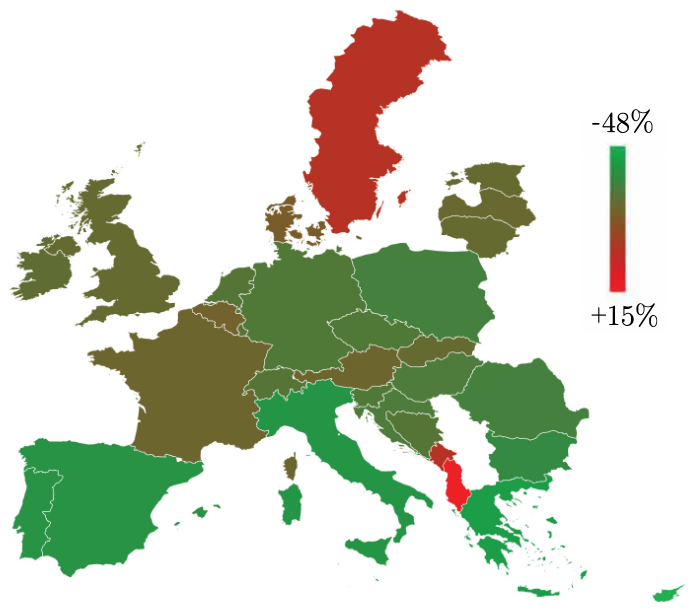}
	\caption{Map of the potential \ac{GHGs} reduction owing to vehicle integrated photovoltaic systems for different regions in Europe~\cite{Scarlat2022QuantificationEurope}.}
	\label{fig:europapa}
\end{figure}

In Fig.~\ref{fig:europapa} we display the impact of the \ac{VIPV} system in different European countries for a fixed vehicle lifetime of 150.000 km.
Specifically, we compare conventional \ac{BEVs} and solar vehicles, whereby both vehicles are optimally designed, minimizing their energy consumption.
The map clearly shows the combined influence of solar irradiance and the electricity mix of the country.
In fact, operating the \ac{VIPV} technology in countries with high solar irradiance (e.g., Greece, Italy, Portugal, Spain) is generally beneficial.
However, if the country already has a low-carbon electricity mix, the additional emissions generated to manufacture the panel may not be compensated by the savings during the operational life, resulting in solar cars being not advantageous (e.g., France), or even detrimental (e.g., Sweden, Montenegro, Albania).
This result underlines the critical importance of the electricity mix at the manufacturing location in addition to the country where the solar vehicle is used. 
Hence, the initial manufacturing emissions influence how beneficial it is to adopt \ac{VIPV} systems, compared to relying on the local grid with \ac{BEVs}.
It is also possible to observe that the adoption of \ac{VIPV} systems always entails an emission shift from the transportation sector in the country of use to the industrial compartment of the country of manufacture, no matter the electricity mix employed in the countries.
This emission delocalization also has ethical implications, as richer countries could yield their emissions to poorer ones by acquiring products manufactured there, paying their way out of environmental responsibilities.
Furthermore, we conducted a sensitivity analysis based on the panel size and the vehicle lifetime.
A larger panel provides more energy but entails more initial emissions at the same time.
Hence, the amount of \ac{GHGs} produced during the panel manufacturing determines a lifetime threshold after which the adoption of \ac{VIPV} becomes advantageous, namely when the cumulative energy saved with the solar panel overcomes the initial manufacturing cost.
Since we assume that both the initial manufacture emissions and power generated scale linearly with the panel area, this tipping line is independent of the panel size and it is only determined by the initial manufacturing cost and the local electricity mix. 
While in countries with a dark electricity mix (Fig.~\ref{fig:NLc}), this line is crossed after a short lifetime of the vehicle (not visible in the plot), in countries with a clean electricity mix or low average irradiation index (Fig.~\ref{fig:SWc}) this line could fall after the average lifespan of a vehicle, constituting a deal-breaker for \ac{VIPV} adoption.
%For this reason, we could regard this parameter as a good metric for how beneficial the adoption of this technology is in each country.
Before and after this line, increasing the panel size amplifies the detrimental or beneficial effect, respectively.
Once crossed, the difference in \ac{GHGs} between the \ac{VIPV}s and \ac{BEVs} asymptotically converges to the value of electricity saved during operations, as a function of the panel size.

%% file: Chapters/Conclusion.tex
\section{Conclusion}\label{sec: Conclusion}

This paper presented a framework to estimate the emissions generated by vehicle-integrated photovoltaic systems.
We considered the different processes to manufacture a mono-crystalline silicon panel, from the raw material extraction to the final panel assembly, leveraging a cradle-to-gate life cycle assessment.
Then we accounted for the emissions generated during the operations thanks to a vehicle energy consumption model, opportunely modified to include the solar panel.
We showcased our framework in a realistic case study, whereby the production takes place in China and the final panel assembly in The Netherlands, estimating the manufacturing emissions of 1 m$^2$ of solar panel to 118 kg of $\mathrm{CO_2}$ equivalents.
Furthermore, we compared the lifetime emissions for solar electric vehicles with conventional battery electric vehicles (BEVs) operated in different European countries.
We show that while in general it is beneficial to operate vehicle-integrated photovoltaic systems in countries with a high irradiation index, if the local electricity mix has already a very low carbon intensity, then the additional emissions to manufacture the panel are unnecessary and a longer vehicle lifetime is required to have an advantageous emission balance.

This work opens the field for the following following areas of research: 
First, we want to account for manufacturing and operation-related emissions directly inside the optimization routine instead of minimizing the vehicle's energy consumption. 
Second, we aim to include other powertrain component manufacturing emissions, as the difference in battery and motor sizing could impact the results.
Third, the framework could be extended by including a cost model, to analyze the economic trade-off and the cost of reducing emissions.
Finally, the analysis could benefit from the application of learning curves to analyze the impact of improvements in the technology and electricity mix in the comparison between BEVS and VIPVs.